\documentclass[iop,apjl,numberedappendix]{emulateapj}
\usepackage{natbib}
\usepackage{graphicx}
\usepackage{amsmath,amsfonts}
\usepackage{booktabs}
\usepackage{hyperref}

\hypersetup{colorlinks=false,pdfborder={0 0 0}}

\slugcomment{Accepted for publication in ApJ Letters, February 18, 2016}

\setcitestyle{notesep={; }}

\def\lsim{\mathrel{\raise.3ex\hbox{$<$\kern-.75em\lower1ex\hbox{$\sim$}}}}
\def\gsim{\mathrel{\raise.3ex\hbox{$>$\kern-.75em\lower1ex\hbox{$\sim$}}}}
\def\gtwid{\mathrel{\raise.3ex\hbox{$>$\kern-.75em\lower1ex\hbox{$\sim$}}}}
\def\proptwid{\mathrel{\raise.3ex\hbox{$\propto$\kern-.75em\lower1ex\hbox{$\sim$}}}}

\begin{document}

\title{ Extreme Brightness Temperatures and Refractive Substructure in 3C\,273 with \textit{RadioAstron} }
\shorttitle{Refractive Substructure in 3C\,273}

\author{Michael D.~Johnson\altaffilmark{1}, Yuri~Y.~Kovalev\altaffilmark{2,3}, Carl R.~Gwinn\altaffilmark{4}, Leonid I.~Gurvits\altaffilmark{5,6}, Ramesh Narayan\altaffilmark{1}, Jean-Pierre Macquart\altaffilmark{7,8}, David L.~Jauncey\altaffilmark{9,10}, Peter~A.~Voitsik\altaffilmark{2}, James M.~Anderson\altaffilmark{12,3}, Kirill V.~Sokolovsky\altaffilmark{2,11}, and Mikhail M.~Lisakov\altaffilmark{2}}
\shortauthors{Johnson et al.}
\altaffiltext{1}{Harvard-Smithsonian Center for Astrophysics, 60 Garden Street, Cambridge, MA 02138, USA}
\altaffiltext{2}{Astro Space Center of Lebedev Physical Institute, Profsoyuznaya 84/32, 117997 Moscow, Russia}
\altaffiltext{3}{Max-Planck-Institute for Radio Astronomy, Auf dem H\"{u}gel 69, D-53121, Germany}
\altaffiltext{4}{Department of Physics, University of California, Santa Barbara, CA 93106, USA}
\altaffiltext{5}{Joint Institute for VLBI ERIC, P.O.\ Box 2, 7990 AA Dwingeloo, The Netherlands}
\altaffiltext{6}{Department of Astrodynamics \& Space Missions, Delft University of Technology, 2629 HS Delft, Delft, The Netherlands}
\altaffiltext{7}{ICRAR/Curtin University, Curtin Institute of Radio Astronomy, Perth, WA 6845, Australia}
\altaffiltext{8}{ARC Centre of Excellence for All-Sky Astrophysics (CAASTRO), Australia}
\altaffiltext{9}{CSIRO Astronomy and Space Sciences, Epping, NSW 1710, Australia, Australia}
\altaffiltext{10}{Research School of Astronomy and Astrophysics, Australian National University, Canberra, ACT, 2611, Australia}
\altaffiltext{11}{Sternberg Astronomical Institute, Moscow State University, Universitetskii~pr.\ 13, 119992 Moscow, Russia}
\altaffiltext{12}{Helmholtz-Zentrum Potsdam, Deutsches GeoForschungsZentrum GFZ, Department 1: Geodesy, Telegrafenberg, 14473, Potsdam, Germany}
\email{mjohnson@cfa.harvard.edu} 


\begin{abstract}
Earth--space interferometry with \textit{RadioAstron} provides the highest direct angular resolution ever achieved in astronomy at any wavelength. \textit{RadioAstron} detections of the classic quasar 3C\,273 on interferometric baselines up to 171\,000~km suggest brightness temperatures exceeding expected limits from the ``inverse-Compton catastrophe'' by two orders of magnitude. We show that at 18~cm, these estimates most probably arise from refractive substructure introduced by scattering in the interstellar medium. We use the scattering properties to estimate an intrinsic brightness temperature of $7{\times}10^{12}~{\rm K}$, which is consistent with expected theoretical limits, but which is ${\sim}15$ times lower than estimates that neglect substructure. At 6.2~cm, the substructure influences the measured values appreciably but gives an estimated brightness temperature that is comparable to models that do not account for the substructure. At $1.35~{\rm cm}$, the substructure does not affect the extremely high inferred brightness temperatures, in excess of $10^{13}~{\rm K}$. We also demonstrate that for a source having a Gaussian surface brightness profile, a single long-baseline estimate of refractive substructure determines an absolute minimum brightness temperature, if the scattering properties along a given line of sight are known, and that this minimum accurately approximates the apparent brightness temperature over a wide range of total flux densities. 
\end{abstract}

\section{Introduction}

Earth--space interferometry with \textit{RadioAstron} extends available interferometric baselines up to ${\sim}360\,000~{\rm km}$, offering unprecedented angular resolution at radio wavelengths and new techniques to study emission processes of the most compact active galactic nuclei (AGN) via direct imaging \citep{Kardashev_2013}. In particular, because the radio emission from AGN is thought to be incoherent synchrotron radiation, the maximum intrinsic brightness temperature is expected to be $\lsim 10^{10.5}~{\rm K}$ if there is equipartition of energy among the particles and fields \citep{Readhead_1994}, or $T_{\rm b} < 10^{11-12}~{\rm K}$ if the particle energies greatly exceed the field energies, set by the ``inverse Compton catastrophe'' \citep{Kellermann_1969}. Even the angular resolution of very-long-baseline interferometry (VLBI) with stations spanning the globe is insufficient to identify brightness temperatures that significantly violate these expected limits.   

VLBI with \textit{RadioAstron} is also sensitive to effects from scattering in the ionized interstellar medium (ISM) that are not detectable on shorter baselines. 
When averaged over long timescales -- days to months -- the scattering blurs compact features in the image, resulting in lower apparent brightness temperatures. However, on shorter timescales, including individual observing epochs, scattering exaggerates image gradients and introduces spurious compact features, or ``refractive substructure,'' within the scattered image \citep{Johnson_Gwinn_2015}. 

In this letter, we study implications of refractive substructure for \textit{RadioAstron} observations of the bright, nearby, $z\approx 0.158$, quasar 3C\,273 at wavelengths of $\lambda=18$, 6.2, and 1.35~cm reported by \citet{Kovalev_2015}.

\section{Scattering Theory}
\label{sec::Scattering}

\subsection{Interstellar Scattering}

Density inhomogeneities in the ionized ISM scatter radio waves, causing scintillation of compact sources, temporal broadening of sharp pulses, and angular broadening of images. The inhomogeneities are often well-described as being localized to a thin ``screen'' between the observer and the source, with a Kolmogorov-like turbulent cascade on scales ranging from ${\gsim}100~{\rm AU}$ to ${\lsim}1000~{\rm km}$ \citep{Armstrong_1995}. The three-dimensional power spectrum of the density fluctuations is $P(\mathbf{q}) \sim C_{\rm N}^2 |\mathbf{q}|^{-(\alpha+2)}$, where $\alpha=5/3$ for Kolmogorov turbulence. 
For reviews of interstellar scattering and scintillation, see \citet{Rickett_1990} or \citet{Narayan_1992}; \citet{Gurvits_1993}, \citet{Shishov_2006}, and \citet{Koay_Macquart_2015} specifically address some of the effects of scattering for space VLBI.

This simple but effective scattering model then depends on two characteristic length scales in addition to the power-law exponent $\alpha$:
\begin{enumerate}
 \item The phase coherence length, $r_0 \propto \lambda^{-2/\alpha}$, is given by the lateral displacement on the scattering screen over which the root-mean-square difference in the random scattering phase is one radian.
\item The Fresnel scale, $r_{\rm F} = \sqrt{ \frac{D R}{D+R}\frac{\lambda}{2\pi}}$, depends on the distances $D$, from the observer to the scattering material, and $R$, from the source to the scattering material. It gives the lateral displacement at which the extra path length relative to the direct path introduces a half radian of phase. 
 \end{enumerate}
 
At $\lambda=18~{\rm cm}$, the scattering along most lines of sight through the Galaxy is ``strong'': $r_0 \ll r_{\rm F}$  \citep{Walker_1998}. In this regime, the scattered image of a point source extends over a few times the refractive scale, $r_{\rm R} = r_{\rm F}^2/r_0 \propto \lambda^{1+2/\alpha}$, and so the stochastic phase introduced by scattering varies by many turns across the scattered image. Note that $r_{\rm R}$ is defined as a lateral length scale on the scattering screen, at a distance $D$ from the observer. At $\lambda=1.35~{\rm cm}$, the scattering along most lines of sight through the Galaxy is ``weak'': $r_0 \gg r_{\rm F}$. In this regime, the scattered image extends over a few times the Fresnel scale, and so the stochastic phase introduced by scattering varies by less than a radian across the scattered image.  

Historically for VLBI, the telltale signature of strong scattering has been smearing of images with a scattering kernel of size $r_{\rm R} \proptwid \lambda^{2}$, resulting in \emph{lower} apparent brightness temperatures. 
This effect of scattering is an ensemble-average property -- it assumes averaging a scattered image over an infinite time and is therefore deterministic \citep{Fish_2014}.

When averaged over shorter timescales, the scattering imparts stochastic signatures, collectively referred to as ``scintillation.'' Point sources such as pulsars commonly show dramatic 100\% modulation of intensity in frequency and time from ``diffractive'' scintillation. The sizes of larger sources, such as AGN, typically quench the diffractive scintillation, just as stars twinkle but planets do not, and the degree to which the scintillation is quenched can provide estimates of intrinsic source size \citep[see, e.g.,][]{Readhead_Hewish_1972,Narayan_1992,Gwinn_1998}. 
However, ``refractive'' scattering effects, reflecting fluctuations on larger scales, can persist for AGN.  

Refractive effects are wideband and typically evolve on timescales of days to weeks, although in some sources the timescale can be as short as hours or less, e.g., PKS\,0405-385 \citep{Kedziora-Chudczer_1997}. Moreover, while diffractive effects such as scatter broadening become weaker at higher frequencies in the strong-scattering regime, refractive effects become stronger until the transition to weak scattering. The most familiar example of refractive scintillation for AGN is flux modulation \citep{RCB_1984}, and monitoring programs have now systematically studied the flux modulation of hundreds of AGN at frequencies of ${\sim}2{-}8~{\rm GHz}$ \citep[e.g.,][]{Rickett_2006,Lovell_2008}.

\subsection{Refractive Substructure}
\label{sec::Substructure}

\citet{NarayanGoodman89} and \citet{GoodmanNarayan89} discovered another effect from refractive scattering: substructure within the scattered image of a point source. \citet{Johnson_Gwinn_2015} showed that this substructure would persist for an extended source, even producing features on angular scales much finer than those intrinsic to the source. 
Refractive substructure is most easily understood in the geometrical optics limit, where scattering ``shuffles'' brightness elements of the image, with individual elements being magnified and demagnified across the image but with unchanged brightness. 

Refractive substructure thereby produces fluctuations -- ``refractive noise'' -- in measured (complex) interferometric visibilities. The effects are especially apparent on long baselines, which are sensitive to the introduced power at small angular scales \citep[see, e.g.,][]{Gwinn_2014}. In the strong scattering regime, an approximate expression for the root-mean-square fluctuations from substructure on a long\footnote{Specifically, the baseline must be long enough to resolve the ensemble-average image $\theta_{\rm img}$.} baseline is \citep{Johnson_Gwinn_2015,GoodmanNarayan89}
\begin{align}
\sigma_{\rm ref} &\approx \sqrt{\frac{\Gamma(4/\alpha)}{2^{2-\alpha}}\frac{\Gamma\left(1+\frac{\alpha}{2}\right)}{\Gamma\left(1-\frac{\alpha}{2}\right)}} \left( \frac{r_0}{r_{\rm F}} \right)^{2-\alpha}\\
\nonumber &\qquad \qquad \times \left(\frac{B}{\left(1+\frac{D}{R}\right)r_0}\right)^{-\frac{\alpha}{2}} \left( \frac{\theta_{\rm scatt}}{\theta_{\rm img}}\right)^2.
\end{align}
Here, $\theta_{\rm scatt} \approx \frac{\sqrt{2\ln{2}}}{\pi}\frac{\lambda}{(1+D/R)r_0}$ is the scattered angular size of a point source, $\theta_{\rm src}$ is the intrinsic angular size of the source, and $\theta_{\rm img} \approx \sqrt{\theta_{\rm src}^2 + \theta_{\rm scatt}^2}$ is the ensemble-average angular size (i.e., the scatter-broadened size). Throughout this paper, we use the full width at half maximum (FWHM) to define the angular size of a source. 
Note that $\sigma_{\rm ref}$ is dimensionless and quantifies the fluctuations as a fraction of the total compact flux density, $F_0$. Eq.~32 from  \citet{Johnson_Gwinn_2015} provides a more general expression that is accurate for arbitrary source structure, on all baselines, and in both the weak and strong scattering regimes but which requires numerical integration to evaluate. When substructure dominates the signal on a baseline, the interferometric visibilities will be zero-mean, complex Gaussian random variables with standard deviation $\sigma_{\rm ref}$, so visibility amplitudes will be drawn from a Rayleigh distribution. The refractive noise will be correlated over a timescale of ${\sim}D\theta_{\rm img}/V_{\perp}$, where $V_{\perp}$ is the characteristic relative transverse velocity of the Earth and scattering material (throughout this paper, we will use $V_{\perp} = 50~{\rm km/s}$ to estimate timescales). The refractive noise in interferometric visibilities will also be correlated among different interferometric baselines $\{u,\,v\}$ with a correlation scale of ${\sim}\lambda/\theta_{\rm img} \sim r_0 \theta_{\rm scatt}/\theta_{\rm img}$. See \S3.2 of \citet{Johnson_Gwinn_2015} for a more complete description and derivation of these correlation scales. 

For comparison with observations of AGN at wavelengths shorter than 20 cm, we consider the case in which the typical scattering angle is smaller than the intrinsic source size. In this case, $\theta_{\rm img} \approx \theta_{\rm src}$, so $\sigma_{\rm ref} \propto B^{-\alpha/2} \lambda^{2+\alpha/2} \theta_{\rm src}^{-2}$. For a source with the same brightness temperature at all wavelengths, $\theta_{\rm src}\propto \lambda$; this is the typical scaling for a self-absorbed jet \citep{Blandford_Konigl_1979}. In this case, the refractive noise on a given baseline increases with wavelength, $\sigma_{\rm ref}\propto \lambda^{\alpha/2} \proptwid \lambda^{5/6}$, because although refractive effects become stronger at shorter wavelengths as noted above, quenching of the refractive scintillation by finite source size dominates the wavelength scaling.

In short, refractive scattering exaggerates image gradients and introduces small-scale features into images. When interpreted in the context of a smooth source model, these features may suggest high brightness temperatures, although a perfect image reconstruction would show no brightness temperatures higher than those of the unscattered source. 

\subsection{Minimum Brightness Temperature Inferred from Substructure}
\label{sec::Tb_min}

When estimating the brightness temperature with sparse baseline coverage, images cannot be reliably formed so the total flux density $F_0$ of a compact component may not be securely estimated (for simplicity, we will henceforth refer to $F_0$ as the ``core'' flux density). Nevertheless, for an assumed Gaussian source with a central brightness temperature $T_{\rm b}$, a single estimate of the correlated flux density $F_{B}$ on a long baseline $B$ determines an absolute minimum for the apparent brightness temperature, even without knowledge of $F_0$ \citep{Lobanov_2015}: 
\begin{align}
\label{eq::Tb_min_intrinsic}
T_{\rm b, min} \approx 6.18 \times 10^{11}~{\rm K} \left( \frac{ B}{{10^5}\,{\rm km}} \right)^2 \left( \frac{F_{B}}{\rm 20~mJy} \right).
\end{align}
This minimum brightness temperature is achieved when $F_{B}/F_0 = 1/e$. Note that this estimate does not correct for scatter-broadening.

When the scattering properties along a given line of sight can be estimated a priori, a single long-baseline measurement that is dominated by refractive noise likewise determines a lower limit on brightness temperature, even when $F_0$ is not known. 
This lower limit occurs when the source becomes large so that the refractive noise is quenched as $\sigma_{\rm ref} \propto \theta_{\rm src}^{-2}$ (see \S\ref{sec::Substructure}). 
When refractive noise is dominant, $F_{B}$ will be drawn from a Rayleigh distribution, so the maximum-likelihood estimator of $\sigma_{\rm ref}$ is $F_{B}/F_0$.  
This then implies that $\theta_{\rm src} \propto \sqrt{F_0}$ and so $T_{\rm b} \propto F_0/\theta_{\rm src}^2 = {\rm constant}$. Thus, the inferred brightness temperature asymptotes to a constant value as $F_{B}/F_0$ becomes small. Taking the Kolmogorov scattering index $\alpha=5/3$ and scaling with median values of the galactic scattering parameters at $\lambda=18~{\rm cm}$ \citep[see][]{Johnson_Gwinn_2015}, we obtain
\begin{align}
\label{eq::Tb_min_scattering}
\nonumber T_{\rm b,min} &= 1.2 \times 10^{12}~{\rm K} \left( \frac{B}{10^5~{\rm km}} \right)^{5/6} \left( \frac{F_{B}}{20~{\rm mJy}} \right) \\ 
& \quad \times \left( \frac{ D }{ 1~{\rm kpc}} \right)^{1/6}  \left( \frac{\lambda}{18~{\rm cm}} \right) \left( \frac{\theta_{\rm scatt}}{300~\mu{\rm as}} \right)^{-5/6}\!\!\!.
\end{align}
Note that this estimate corrects for scatter-broadening.

\begin{figure}[t]
\centering
\includegraphics[width=0.47\textwidth]{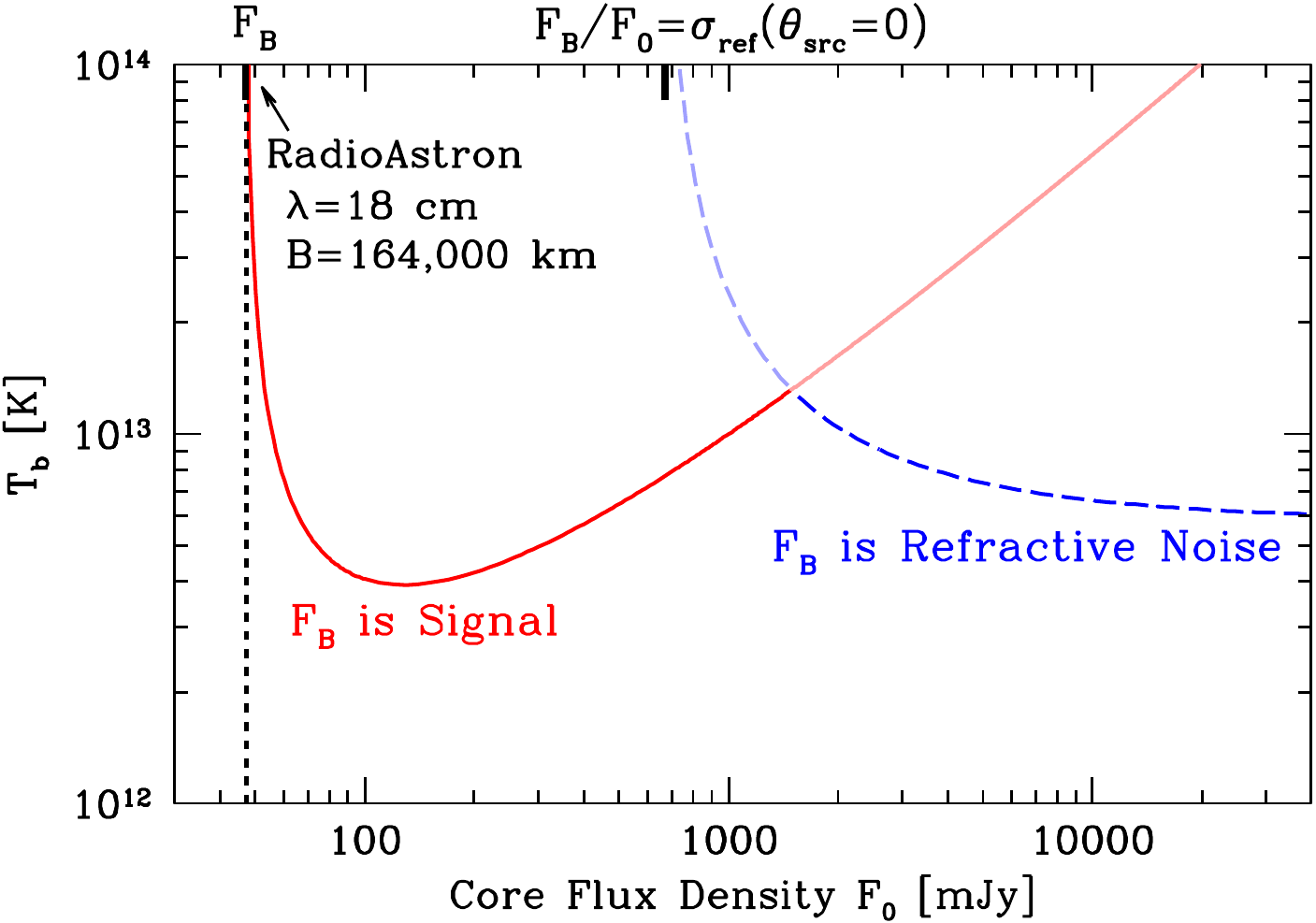}
\caption
{ 
Inferred brightness temperature as a function of the core flux density, $F_0$, using the average correlated flux density for \textit{RadioAstron} detections at $\lambda=18~{\rm cm}$ ($F_{B} = 47~{\rm mJy}$ on a baseline $B=164\,000~{\rm km}$). The red solid line shows the inferred brightness temperature for a Gaussian model with no scattering (Eq.~\ref{eq::Tb_min_intrinsic}); the blue dashed line shows the inferred brightness temperature if the \textit{RadioAstron} detections are refractive noise (Eq.~\ref{eq::Tb_min_scattering}). The lower of the two curves determines whether intrinsic structure or scattering dominates $F_{B}$. 
$T_{\rm b}$ from intrinsic structure diverges when the core is completely unresolved ($F_{B} = F_0$), while $T_{\rm b}$ from refractive substructure diverges when the normalized visibility $F_{B}/F_0$ is equal to the expected refractive noise for a point source. For any core flux density $F_0$ greater than ${\sim}1~{\rm Jy}$, the measurements will be dominated by refractive noise with an inferred $T_{\rm b} \approx 7\times 10^{12}~{\rm K}$ that only weakly depends on the unknown core flux density. For 3C\,273, $F_0 \approx 5.0~{\rm Jy}$. 
}
\label{fig_Tb_vs_F0}
\end{figure}

Figure~\ref{fig_Tb_vs_F0} compares the inferred brightness temperatures from intrinsic structure and from refractive substructure for the \textit{RadioAstron} 18~cm observations (described below) as a function of the core flux density, $F_0$. 

\begin{figure*}[t]
\centering
\includegraphics*[width=1.0\textwidth]{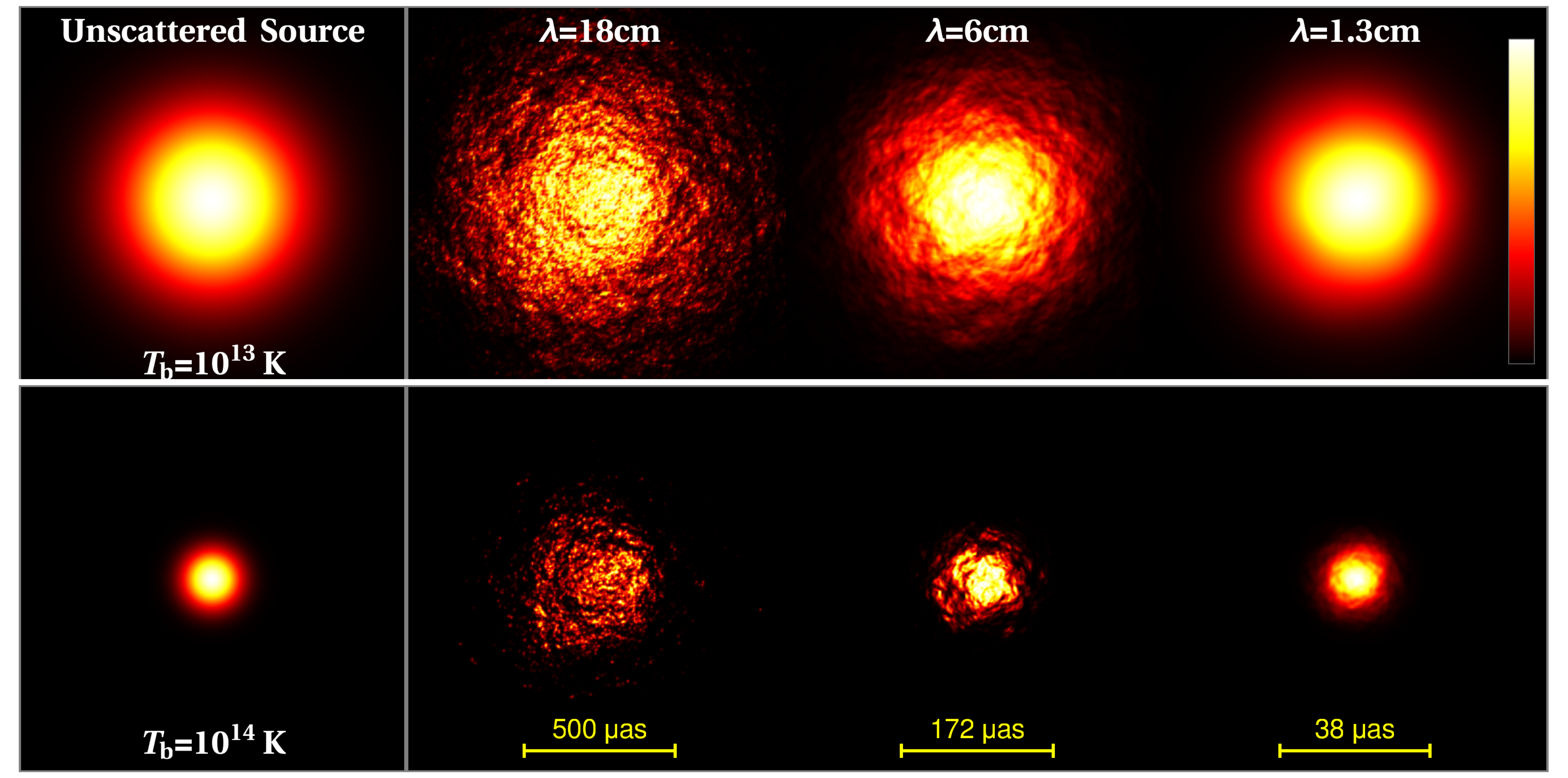}
\caption
{ 
Simulated images showing the effects of refractive substructure at $\lambda=18$, 6.2, and 1.35~cm. For each wavelength two cases are shown, peak brightness temperatures of ({\it top}) $T_{\rm b} = 10^{13}~{\rm K}$ and ({\it bottom}) $T_{\rm b} = 10^{14}~{\rm K}$. To simplify the comparison, a 5~Jy circular Gaussian intrinsic source is assumed in all cases. Brightness is shown on a linear scale and is scaled so that the maximum image brightness is identical across each panel. The angular range in each panel is scaled linearly with wavelength so that the unscattered source would appear identical across each panel (hence, a fixed physical observing array would have the same beam size across each panel horizontally) with an equal angular scale in the top and bottom panels. The scattering parameters correspond to the NE2001 estimates for 3C\,273 (see \S\ref{sec::Analysis}). The effects of substructure at 18 and 6~cm are readily apparent even when the typical scattering angle is smaller than the intrinsic angular structure. 
}
\label{fig_Scattering_Example}
\end{figure*}

\section{Observations}
\label{sec::Observations}

The observations of 3C\,273 reported by \citet{Kovalev_2015} were made in December 2012 to February 2013 with \textit{RadioAstron} in concert with the Green Bank Telescope (GBT), the phased Karl G.~Jansky Very Large Array (VLA), the 100-meter Effelsberg radio telescope, and the 305-meter Arecibo Telescope at $\lambda{=}18$, 6.2, and 1.35~cm. 3C\,273 was detected on baselines exceeding $100\,000~{\rm km}$ at each of these wavelengths. 
At 18~cm, detections on two epochs had correlated flux densities of $42\pm 7~{\rm mJy}$ and $52\pm 9~{\rm mJy}$ on baselines of $157\,000~{\rm km}$ and $171\,000~{\rm km}$, respectively. At 6.2~cm, detections on two epochs had correlated flux densities of $125\pm 17~{\rm mJy}$ and $123\pm 19~{\rm mJy}$ on baselines of $90\,000~{\rm km}$ and $103\,000~{\rm km}$, respectively. At 1.35~cm, the single epoch with a detection found a correlated flux density of $125\pm 22~{\rm mJy}$ on a baseline of $103\,000~{\rm km}$. These errors include amplitude calibration uncertainties; the fringe amplitude had a signal-to-noise ratio ${\gsim}10$ in all cases. \citet{Kovalev_2015} estimate the core flux densities to be $F_0 = 5.0~{\rm Jy}$ at 18~cm, $4.3~{\rm Jy}$ at 6.2~cm, and $3.4~{\rm Jy}$ at 1.35~cm.

\section{Scattering and Substructure of 3C\,273}
\label{sec::Analysis}

\begin{figure*}[t]
\centering
\includegraphics*[width=\textwidth]{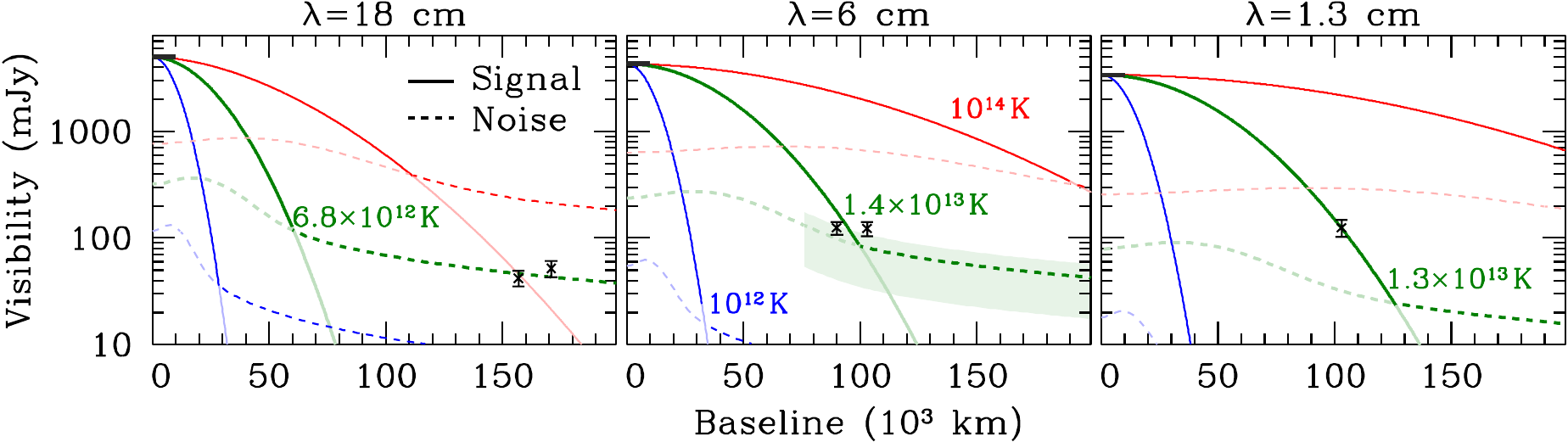}
\caption
{ Visibility amplitude vs.\ baseline length for circular Gaussian source models; the solid line shows the ensemble-average visibility amplitude (``signal'') while the dashed line shows the root-mean-square refractive fluctuations $\sigma_{\rm ref}$ (``noise''). Each model has a total compact flux density that is equal to the estimates that \citet{Kovalev_2015} derived via an imaging analysis with ground-based VLBI; the corresponding zero-baseline value and sampled baseline lengths (up to ${\sim}10^4~{\rm km}$) are denoted by a heavy horizontal tick in each panel.  
At each baseline, the higher of the two curves determines whether intrinsic structure or refractive noise will dominate measured visibilities. \textit{RadioAstron} detections are shown as black points with ${\pm}1\sigma$ error bars that include calibration uncertainties.  Three Gaussian models are also shown: brightness temperatures of $10^{12}~{\rm K}$ (blue), $10^{14}~{\rm K}$ (red), and wavelength-dependent best-fit models to the \textit{RadioAstron} long-baseline detections (green). At $\lambda{=}18~{\rm cm}$, the \textit{RadioAstron} detections are dominated by refractive noise, and the noise level provides an estimate of the apparent brightness temperature; at $6.2~{\rm cm}$, the detections reflect a combination of ensemble-average visibility and refractive noise; and at $1.35~{\rm cm}$, the detections are dominated by the ensemble-average visibility. The center figure also shows the middle 68\% range of refractive noise amplitudes expected for the middle model as a shaded region. The range is highly asymmetric -- within individual epochs, null detections are much more likely than amplitudes significantly higher than the root-mean-square noise. Note also that the zero-baseline noise for each model gives the predicted refractive modulation of the total flux density for the compact component.
}
\label{fig_Models_wData}
\end{figure*}

Because 3C\,273 lies at galactic coordinates $\ell = 289.95^\circ$ and $b=+64.36^\circ$, the scattering properties along its line of sight are typical of extragalactic sources that are well away from the Galactic plane. Moreover, because angular broadening preferentially weights nearby material \citep{Blandford_Narayan_1985}, the substructure for extragalactic sources is overwhelmingly dominated by scattering within the Milky Way \citep[see, e.g.,][]{Koay_Macquart_2015}.  
For 3C\,273, the NE2001 model of the Galactic distribution of free electrons predicts an angular broadening of $\theta_{\rm scatt} \approx 0.75~{\rm mas}$ at an observing frequency of $1~{\rm GHz}$ \citep{NE2001}. We scale this estimate to other wavelengths assuming that $\theta_{\rm scatt} \propto \lambda^2$ (a Kolmogorov scaling, $\theta_{\rm scatt} \propto \lambda^{11/5}$, gives similar results). Figure~\ref{fig_Scattering_Example} shows simulated images of the refractive scattering in each \textit{RadioAstron} observing band.

At $\lambda=18~{\rm cm}$, detections with \textit{RadioAstron} imply a brightness temperature of $T_{\rm b} \gsim 10^{14}~{\rm K}$ for Gaussian source models with no refractive substructure (after accounting for scatter broadening). However, the expected refractive noise on these baselines would then be ${\sim}5$ times larger than the observed signal (see Figure~\ref{fig_Models_wData}). Thus, these Earth--space visibilities are most likely the first detections of refractive substructure in an extragalactic source. To reproduce the observed long-baseline detections via refractive noise requires a source size of approximately $570~\mu{\rm as}$ with a corresponding brightness temperature of $T_{\rm b} \sim 6.8 \times 10^{12}~{\rm K}$ if the core flux density is $F_0 = 5.0~{\rm Jy}$. This estimate is rather insensitive to the assumed core flux density (see Figure~\ref{fig_Tb_vs_F0}).

At $\lambda=6.2~{\rm cm}$, estimates from intrinsic structure and from substructure both give $T_{\rm b} \sim 1.4 \times 10^{13}~{\rm K}$ (see Figure~\ref{fig_Models_wData}). Because the contribution of refractive noise is stochastic, one could determine whether substructure is dominant by examining the correlated flux density on different observing epochs separated by more than $D\theta_{\rm img}/V_{\perp} \sim 1$~week. Absence of variation would be a secure indication that long-baseline measurements are not the result of substructure.

At $\lambda=1.35~{\rm cm}$, the expected refractive noise for a Gaussian source that matches the long-baseline detections ($\theta_{\rm src} = 26~\mu{\rm as}$) is a factor of ${\sim}4$ smaller than the observed signal, showing that these detections are likely signal-dominated for a source with brightness temperature $T_{\rm b} \approx 1.3 \times 10^{13}~{\rm K}$. The detections also determine an upper limit, $T_{\rm b} \lsim 7 \times 10^{13}~{\rm K}$, for the brightness temperature, since the refractive noise of a significantly more compact source would exceed the measured visibilities. 

Although we have assumed specific parameters to describe the scattering, the inferred brightness temperature is not sensitive to changes in these parameters. For example, the inferred $T_{\rm b}$ varies by only 10\% for $\alpha$ ranging between 1.66 and 1.9 while fixing $\theta_{\rm scatt}$. Likewise, a scattering screen placed at $10~{\rm kpc}$ rather than $1~{\rm kpc}$ would have a corresponding refractive noise that is $10^{-1/6}\approx 0.68$ times the current estimates, so the estimated brightness temperature from substructure would be \emph{higher} by a factor of $10^{1/6} \approx 1.47$. Our estimates of $T_{\rm b}$ also depend on $\theta_{\rm scatt}$ (see, e.g., Eq.~\ref{eq::Tb_min_scattering}), which we have estimated using the NE2001 model. Previous observations have found tolerable agreement with this model, often to within a factor of ${\sim}2$ \citep{Lazio_2008}, although more detailed study is essential and it remains difficult to confidently identify the weak angular broadening of high-latitude sources. 

The stochastic nature of refractive noise also contributes uncertainty to estimates of $T_{\rm b}$. A measured set of long-baseline visibilities determines an estimate of $\sigma_{\rm ref}$, and for measurements that are dominated by refractive noise, $\sigma_{\rm ref} \propto T_{\rm b}$. Because the Rayleigh distribution contains significant power for samples below the root-mean-square, the posterior distribution of $\sigma_{\rm ref}$ may be weakly constrained: for a single measurement, the middle $\pm1\sigma$ range for the posterior distribution of $\sigma_{\rm ref}$ spans a factor of ${\approx}7$, while for two measurements, the inner $\pm 1\sigma$ range spans a factor of ${\approx}2.5$. In each case, the range is highly asymmetric about the root-mean-square, primarily extending to larger values of $\sigma_{\rm ref}$. Thus, the true brightness temperatures may be up to a few times higher than what we have inferred but could not be significantly lower. Conversely, since our measurements only address the detections with \textit{RadioAstron} on long baselines, and there were other observing epochs with no detections on comparable baselines, these measurements may be biased by sampling only the high end of the Rayleigh distribution. In this case, the root-mean-square refractive noise may be up to a factor of ${\sim}2$ smaller than our estimates implying a \emph{lower} $T_{\rm b}$ by a factor of ${\sim}2$.

\section{Summary}
\label{sec::Summary}

On long baselines, refractive noise from interstellar scattering likely dominates \textit{RadioAstron} detections of 3C\,273 at $\lambda{=}18~{\rm cm}$, is probably comparable to the observed signal at $6.2~{\rm cm}$, and is likely insignificant at $1.35~{\rm cm}$. At 18~cm, the brightness temperature estimate after accounting for refractive noise is $T_{\rm b} \sim 7 \times 10^{12}~{\rm K}$. This brightness temperature at 18~cm is 15 times lower than the estimate for a smooth, scatter-broadened Gaussian source. 

Our results, the first detection of refractive substructure in an extragalactic source, demonstrate the importance of refractive substructure for Earth--space VLBI. 
Traditionally, refractive flux modulation has been proposed as the most promising signature of scattering to study in parallel with space VLBI \citep[e.g.,][]{Dennison_1993}. However, studies of flux modulation require regular monitoring of the total and correlated flux, they cannot always unambiguously disentangle the scattering fluctuations from those that are intrinsic to the source, and the results can be sensitive to the unknown fraction of the total flux density in the compact, scintillating component. Note also that the flux modulation predicted for 3C\,273 in our models (see Figure~\ref{fig_Models_wData}) is probably only detectable at 18~cm, and at that wavelength would have a decorrelation timescale $D\theta_{\rm img}/V_{\perp} \sim 1$~month that is comparable to the timescale for intrinsic variability. In contrast, studies of refractive scintillation with Earth--space VLBI can obtain meaningful information about the source and scattering with individual observing epochs, and the results are insensitive to the total flux density. Consequently, estimates of brightness temperature with sparse baseline coverage can be \emph{improved} by refractive scattering. 
If multiple epochs are combined, then additional detailed information can be derived, including the scattering timescales involved. For instance, the dependence of refractive noise on baseline length determines the power-law exponent $\alpha$ for large-scale turbulence in the scattering region, which we have simply assumed to be the Kolmogorov value ($\alpha=5/3$). Thus, our discovery of refractive substructure in AGN offers a new, robust pathway for estimating the brightness temperatures of compact sources with sparse baseline coverage and for studies of large-scale scattering.

\acknowledgements{MDJ thanks the Gordon and Betty Moore Foundation for financial support of this work through grant GBMF-3561 to Sheperd Doeleman. The \textit{RadioAstron} project is led by the Astro Space Center of the Lebedev Physical Institute of the Russian Academy of Sciences and the Lavochkin Association of the Russian Federal Space Agency, and is a collaboration with partner institutions in Russia and other countries. This project was supported by the Russian Foundation for Basic Research
grant 13-02-12103.
The Arecibo Observatory is operated by SRI International under a cooperative
agreement with the National Science Foundation (AST-1100968), and in alliance with Ana G.~Mendez-Universidad Metropolitana, and the Universities Space Research Association. The
National Radio Astronomy Observatory is a facility of the National Science Foundation operated
under cooperative agreement by Associated Universities, Inc. This research is partly based on
observations with the 100-meter telescope of the MPIfR (Max-Planck-Institute for Radio
Astronomy) at Effelsberg. 
}

{\it Facilities:} \facility{RadioAstron Space Radio Telescope (Spektr-R)}, \facility{Arecibo}, \facility{GBT}, \facility{VLA}, \facility{Effelsberg}\\ \ \\


\begin{thebibliography}{}
\expandafter\ifx\csname natexlab\endcsname\relax\def\natexlab#1{#1}\fi

\bibitem[{{Armstrong} {et~al.}(1995){Armstrong}, {Rickett}, \&
  {Spangler}}]{Armstrong_1995}
{Armstrong}, J.~W., {Rickett}, B.~J., \& {Spangler}, S.~R. 1995, \apj, 443, 209

\bibitem[{{Blandford} \& {Narayan}(1985)}]{Blandford_Narayan_1985}
{Blandford}, R., \& {Narayan}, R. 1985, \mnras, 213, 591

\bibitem[{{Blandford} \& {K{\"o}nigl}(1979)}]{Blandford_Konigl_1979}
{Blandford}, R.~D., \& {K{\"o}nigl}, A. 1979, \apj, 232, 34

\bibitem[{{Cordes} \& {Lazio}(2002)}]{NE2001}
{Cordes}, J.~M., \& {Lazio}, T.~J.~W. 2002, arXiv:astro-ph/0207156

\bibitem[{{Dennison} {et~al.}(1993){Dennison}, {Fiedler}, {Johnston}, \&
  {Simon}}]{Dennison_1993}
{Dennison}, B., {Fiedler}, R.~L., {Johnston}, K.~J., \& {Simon}, R.~L. 1993, in
  Propagation Effects in Space VLBI, ed. L.~I. {Gurvits}, 23

\bibitem[{{Fish} {et~al.}(2014){Fish}, {Johnson}, {Lu}, {Doeleman}, {Bouman},
  {Zoran}, {Freeman}, {Psaltis}, {Narayan}, {Pankratius}, {Broderick}, {Gwinn},
  \& {Vertatschitsch}}]{Fish_2014}
{Fish}, V.~L., {Johnson}, M.~D., {Lu}, R.-S., {et~al.} 2014, \apj, 795, 134

\bibitem[{{Goodman} \& {Narayan}(1989)}]{GoodmanNarayan89}
{Goodman}, J., \& {Narayan}, R. 1989, \mnras, 238, 995

\bibitem[{{Gurvits}(1993)}]{Gurvits_1993}
{Gurvits}, L.~I., ed. 1993, {Propagation Effects in Space VLBI}

\bibitem[{{Gwinn} {et~al.}(1998){Gwinn}, {Britton}, {Reynolds}, {Jauncey},
  {King}, {McCulloch}, {Lovell}, \& {Preston}}]{Gwinn_1998}
{Gwinn}, C.~R., {Britton}, M.~C., {Reynolds}, J.~E., {et~al.} 1998, \apj, 505,
  928

\bibitem[{{Gwinn} {et~al.}(2014){Gwinn}, {Kovalev}, {Johnson}, \&
  {Soglasnov}}]{Gwinn_2014}
{Gwinn}, C.~R., {Kovalev}, Y.~Y., {Johnson}, M.~D., \& {Soglasnov}, V.~A. 2014,
  \apjl, 794, L14

\bibitem[{{Johnson} \& {Gwinn}(2015)}]{Johnson_Gwinn_2015}
{Johnson}, M.~D., \& {Gwinn}, C.~R. 2015, \apj, 805, 180

\bibitem[{{Kardashev} {et~al.}(2013){Kardashev}, {Khartov}, {Abramov},
  {Avdeev}, {Alakoz}, {Aleksandrov}, {Ananthakrishnan}, {Andreyanov},
  {Andrianov}, {Antonov}, {Artyukhov}, {Arkhipov}, {Baan}, {Babakin},
  {Babyshkin}, {Bartel'}, {Belousov}, {Belyaev}, {Berulis}, {Burke},
  {Biryukov}, {Bubnov}, {Burgin}, {Busca}, {Bykadorov}, {Bychkova},
  {Vasil'kov}, {Wellington}, {Vinogradov}, {Wietfeldt}, {Voitsik},
  {Gvamichava}, {Girin}, {Gurvits}, {Dagkesamanskii}, {D'Addario},
  {Giovannini}, {Jauncey}, {Dewdney}, {D'yakov}, {Zharov}, {Zhuravlev},
  {Zaslavskii}, {Zakhvatkin}, {Zinov'ev}, {Ilinen}, {Ipatov}, {Kanevskii},
  {Knorin}, {Casse}, {Kellermann}, {Kovalev}, {Kovalev}, {Kovalenko}, {Kogan},
  {Komaev}, {Konovalenko}, {Kopelyanskii}, {Korneev}, {Kostenko}, {Kotik},
  {Kreisman}, {Kukushkin}, {Kulishenko}, {Cooper}, {Kut'kin}, {Cannon},
  {Larionov}, {Lisakov}, {Litvinenko}, {Likhachev}, {Likhacheva}, {Lobanov},
  {Logvinenko}, {Langston}, {McCracken}, {Medvedev}, {Melekhin}, {Menderov},
  {Murphy}, {Mizyakina}, {Mozgovoi}, {Nikolaev}, {Novikov}, {Novikov},
  {Oreshko}, {Pavlenko}, {Pashchenko}, {Ponomarev}, {Popov}, {Pravin-Kumar},
  {Preston}, {Pyshnov}, {Rakhimov}, {Rozhkov}, {Romney}, {Rocha}, {Rudakov},
  {R{\"a}is{\"a}nen}, {Sazankov}, {Sakharov}, {Semenov}, {Serebrennikov},
  {Schilizzi}, {Skulachev}, {Slysh}, {Smirnov}, {Smith}, {Soglasnov},
  {Sokolovskii}, {Sondaar}, {Stepan'yants}, {Turygin}, {Turygin}, {Tuchin},
  {Urpo}, {Fedorchuk}, {Finkel'shtein}, {Fomalont}, {Fejes}, {Fomina},
  {Khapin}, {Tsarevskii}, {Zensus}, {Chuprikov}, {Shatskaya}, {Shapirovskaya},
  {Sheikhet}, {Shirshakov}, {Schmidt}, {Shnyreva}, {Shpilevskii}, {Ekers}, \&
  {Yakimov}}]{Kardashev_2013}
{Kardashev}, N.~S., {Khartov}, V.~V., {Abramov}, V.~V., {et~al.} 2013,
  Astronomy Reports, 57, 153

\bibitem[{{Kedziora-Chudczer} {et~al.}(1997){Kedziora-Chudczer}, {Jauncey},
  {Wieringa}, {Walker}, {Nicolson}, {Reynolds}, \&
  {Tzioumis}}]{Kedziora-Chudczer_1997}
{Kedziora-Chudczer}, L., {Jauncey}, D.~L., {Wieringa}, M.~H., {et~al.} 1997,
  \apjl, 490, L9

\bibitem[{{Kellermann} \& {Pauliny-Toth}(1969)}]{Kellermann_1969}
{Kellermann}, K.~I., \& {Pauliny-Toth}, I.~I.~K. 1969, \apjl, 155, L71

\bibitem[{{Koay} \& {Macquart}(2015)}]{Koay_Macquart_2015}
{Koay}, J.~Y., \& {Macquart}, J.-P. 2015, \mnras, 446, 2370

\bibitem[{{Kovalev} {et~al.}(2016){Kovalev}, {Kardashev}, {Kellermann},
  {Lobanov}, {Johnson}, {Gurvits}, {Voitsik}, {Zensus}, {Anderson}, {Bach},
  {Jauncey}, {Ghigo}, {Ghosh}, {Kraus}, {Kovalev}, {Lisakov}, {Petrov},
  {Romney}, {Salter}, \& {Sokolovsky}}]{Kovalev_2015}
{Kovalev}, Y.~Y., {Kardashev}, N.~S., {Kellermann}, K.~I., {et~al.} 2016, ApJL,
  submitted, arXiv:1601.05806

\bibitem[{{Lazio} {et~al.}(2008){Lazio}, {Ojha}, {Fey}, {Kedziora-Chudczer},
  {Cordes}, {Jauncey}, \& {Lovell}}]{Lazio_2008}
{Lazio}, T.~J.~W., {Ojha}, R., {Fey}, A.~L., {et~al.} 2008, \apj, 672, 115

\bibitem[{{Lobanov}(2015)}]{Lobanov_2015}
{Lobanov}, A. 2015, \aap, 574, A84

\bibitem[{{Lovell} {et~al.}(2008){Lovell}, {Rickett}, {Macquart}, {Jauncey},
  {Bignall}, {Kedziora-Chudczer}, {Ojha}, {Pursimo}, {Dutka}, {Senkbeil}, \&
  {Shabala}}]{Lovell_2008}
{Lovell}, J.~E.~J., {Rickett}, B.~J., {Macquart}, J.-P., {et~al.} 2008, \apj,
  689, 108

\bibitem[{{Narayan}(1992)}]{Narayan_1992}
{Narayan}, R. 1992, Royal Society of London Philosophical Transactions Series
  A, 341, 151

\bibitem[{{Narayan} \& {Goodman}(1989)}]{NarayanGoodman89}
{Narayan}, R., \& {Goodman}, J. 1989, \mnras, 238, 963

\bibitem[{{Readhead}(1994)}]{Readhead_1994}
{Readhead}, A.~C.~S. 1994, \apj, 426, 51

\bibitem[{{Readhead} \& {Hewish}(1972)}]{Readhead_Hewish_1972}
{Readhead}, A.~C.~S., \& {Hewish}, A. 1972, \nat, 236, 440

\bibitem[{{Rickett}(1990)}]{Rickett_1990}
{Rickett}, B.~J. 1990, \araa, 28, 561

\bibitem[{{Rickett} {et~al.}(1984){Rickett}, {Coles}, \& {Bourgois}}]{RCB_1984}
{Rickett}, B.~J., {Coles}, W.~A., \& {Bourgois}, G. 1984, \aap, 134, 390

\bibitem[{{Rickett} {et~al.}(2006){Rickett}, {Lazio}, \&
  {Ghigo}}]{Rickett_2006}
{Rickett}, B.~J., {Lazio}, T.~J.~W., \& {Ghigo}, F.~D. 2006, \apjs, 165, 439

\bibitem[{{Shishov} {et~al.}(2006){Shishov}, {Coles}, {Rickett}, {Bird},
  {Efimov}, {Samoznaev}, {Rudash}, {Chashei}, {Plettemeier}, {Spangler},
  {Tokarev}, {Belov}, {Boiko}, {Komrakov}, {Chau}, {Harmon}, {Sulzer},
  {Kojima}, {Tokumaru}, {Fujiki}, {Janardhan}, {Jackson}, {Hick}, {Buffington},
  {Olyak}, {Fallows}, {Nechaeva}, {Gavrilenko}, {Gorshenkov}, {Alimov},
  {Molotov}, {Pushkarev}, {Shanks}, {Tuccari}, {Lotova}, {Vladimirski},
  {Obridko}, {Gubenko}, {Andreev}, {Stinebring}, {Gwinn}, {Lovell}, {Jauncey},
  {Senkbeil}, {Shabala}, {Bignall}, {Macquart}, {Rickett}, {Kedziora-Chudczer},
  {Smirnova}, {Rickett}, {Malofeev}, {Malov}, {Tyulbashev}, {Jessner},
  {Sieber}, \& {Wielebinski}}]{Shishov_2006}
{Shishov}, V.~I., {Coles}, W.~A., {Rickett}, B.~J., {et~al.} 2006, {The Journal
  of the Eurasian Astronomical Society}, astro-ph/0609517

\bibitem[{{Walker}(1998)}]{Walker_1998}
{Walker}, M.~A. 1998, \mnras, 294, 307

\end{thebibliography}

\end{document}